\documentclass[aps,twocolumn,pra,superscriptaddress,showpacs,tightenlines,floatfix]{revtex4}
\usepackage{amssymb}
\usepackage{amsmath}
\usepackage{graphicx}
\usepackage{epsfig}
\usepackage{subfigure}
\usepackage{amsfonts}

\begin{document}

\title{Entanglement of Two Distinguishable Atoms in a Rectangular Waveguide: Linear Approximation with Single Excitation}
\author{Jing \surname{Li}}
\affiliation{Synergetic Innovation Center for Quantum Effects and Applications, Key Laboratory for Matter Microstructure and Function of Hunan Province, Key Laboratory of Low-Dimensional Quantum Structures and Quantum Control of Ministry of Education, School of Physics and Electronics, Hunan Normal University, Changsha 410081, China}
\author{Lijuan \surname{Hu}}
\affiliation{College of Science, Hunan University of Science and Engineering, Yongzhou 425199, China}
\author{Jing \surname{Lu}}
\affiliation{Synergetic Innovation Center for Quantum Effects and Applications, Key Laboratory for Matter Microstructure and Function of Hunan Province, Key Laboratory of Low-Dimensional Quantum Structures and Quantum Control of Ministry of Education, School of Physics and Electronics, Hunan Normal University, Changsha 410081, China}
\author{Lan \surname{Zhou}}
\thanks{Corresponding author}
\email{zhoulan@hunnu.edu.cn}
\affiliation{Synergetic Innovation Center for Quantum Effects and Applications, Key Laboratory for Matter Microstructure and Function of Hunan Province, Key Laboratory of Low-Dimensional Quantum Structures and Quantum Control of Ministry of Education, School of Physics and Electronics, Hunan Normal University, Changsha 410081, China}

\begin{abstract}
We consider two two-level systems (TLSs) coupled to the vacuum of guided modes confined in
a rectangular waveguide. Two TLSs are fixed at different points in the waveguide and initially
share an excitation. For the energy separation of the TLSs far away from the cutoff frequencies
of transverse modes, two coupled delay-differential equations are obtained for the probability
amplitudes of the TLSs. The effects of the difference of TLSs' energy separations and the inter-TLS
distance on the time evolution of the concurrence of the TLSs are examined.
\end{abstract}

\pacs{03.65.Yz, 03.65.-w}
%03.65.Yz 	Decoherence; open systems;
%03.65.-w 	Quantum mechanics

\maketitle

\section{Introduction}

The quantum superposition principle allows a system composed of multipartite
quantum systems to has states that cannot be factorized in products of
states of the individual quantum systems. This nonseparability, labeled as
entanglement, is an important physical resource for applications of quantum
information processing. Scalable quantum information processing in quantum
computation and communication is essentially based on a quantum network~\cite%
{Kimble}. A quantum network consists of quantum channels and nodes.
Two-level systems (TLSs) fixed at quantum nodes are called stationary qubits
which generate, store, and process quantum information. It is essentially
important to generate or keep the correlation among TLSs located at
different positions for protecting quantum information. The bipartite
entanglement involving two TLSs is of special interest. Spatially separated
TLSs talking to each other can be either mediated or destroyed via
electromagnetic fields\cite{Dicke,Lehmberg70,Milonni74,Cook35}. Complete
disentanglement is achieved in finite time for two TLSs coupled individually
to two vacuum cavities~\cite{YuESD}. The entanglement exhibits revivals in
time for two TLSs coupled collectively to a multimode vacuum field in free
space~\cite{FicekPRA74}.

To build large scale quantum networks, an electromagnetic field in a
one-dimensional (1D) waveguide is of special interest. The electromagnetic
field is confined spatially in two dimensions and propagates along the
remaining one, so it consists of infinite modes for right and left-going
photons of continuous varying frequencies. Spontaneously emitted waves from
the TLS will interfere with the incident wave~\cite%
{ZLQrouter,Fans,ZLPRL08,DongPRA,Zheng,LawPRA78,TShiSun}. The coupling of the
electromagnetic field to a TLS can be increased by reducing the transverse
size. A waveguide with a cross section has many guided modes, e.g.
transverse-magnetic (TM) modes or transverse-electric (TE) ones. However,
most work only consider TLSs interacting with one guided mode of the
waveguide~\cite%
{Fans,ZLPRL08,DongPRA,Zheng,LawPRA78,TShiSun,LaaPRL113,PRA14Red,Ordonez,FangPRA91,RoyRMP89,KanuPRL124,KanuarXiv2006}%
, which means that the transverse-size effect has been ignored. In this
paper, we study the time evolution of entanglement measure for two uncoupled
TLSs interacting with the electromagnetic field confined in a 1D rectangular
hollow metallic waveguide. The TLSs share initially an excitation and the
field is in vacuum. Such waveguide has many guided modes. There is a
continuous range of frequencies and a minimum frequency (called cutoff
frequency) allowed in each guided mode~\cite{TETMmode}. when the transitions
of the TLSs are far away from the cutoff frequencies of guided modes, the
probability amplitudes the TLSs is described by the delay differential
equations by tracing out the continuum of bosonic modes in the waveguide.
The spatial separation of the two TLSs introduces the position-dependent
phase factor and the time delay (finite time required for light to travel
from one TLS to the other) in each transverse mode. The phase factors and
the time delays are different in different transverse modes. The effect of
the phase factors and the time delays on the entanglement dynamics of the
TLSs are studied in details by considering the TLSs interacting with single
transverse mode and double transverse modes.

This paper is organized as follows. In Sec.~\ref{Sec:2}, we introduce the
model and establish the notation. In Sec.~\ref{Sec:3}, we derive the
relevant equations describing the dynamics of the system for the case of the
TLSs being initially sharing an excitation and the waveguide mode in the
vacuum state. In Sec.~\ref{Sec:4}, we analyze the behavior of the TLSs'
concurrence when the TLSs interact resonantly with the electromagnetic field
of one or two guided modes. We make a conclusion in Sec.~\ref{Sec:5}.

%%%%%%%%%%%%%%%%%%%%%%%%%%%%%%%%%%%%%%%%%%%%%%%%%%%%%%%%%%%%

\section{\label{Sec:2}Two TLSs in a rectangular waveguide}

%%%%%%%%%%%%%%%%%%%%%%%%%%%%%%%%%%%%%%%%%%%%%%%%%%%%%%%%%%%%
The Hamiltonian of the TLSs interacting with the electromagnetic field of a
rectangular waveguide consist of three parts
\begin{equation}
\hat{H}=\hat{H}_{a}+\hat{H}_{f}+\hat{H}_{int}.  \label{2-A1}
\end{equation}%
The fist part is the free Hamiltonian of the TLSs
\begin{equation}
\hat{H}_{a}=\sum\limits_{l=1}^{2}\hbar \omega _{l}\hat{S}_{l}^{+}\hat{S}_{l}^{-},  \label{2-A2}
\end{equation}%
where $\hat{S}_{l}^{+}\equiv \left\vert e_{l}\right\rangle \left\langle
g_{l}\right\vert $ ($\hat{S}_{l}^{-}\equiv \left\vert g_{l}\right\rangle
\left\langle e_{l}\right\vert $ ) is the rising (lowing) atomic operator of
the $l$-$th$ TLS, $\omega _{l}(l=1,2)$ are the energy difference between the
excited state $|e\rangle $ and the ground state $|g\rangle $.
\begin{figure}[tbp]
\includegraphics[clip=true,height=6cm,width=8cm]{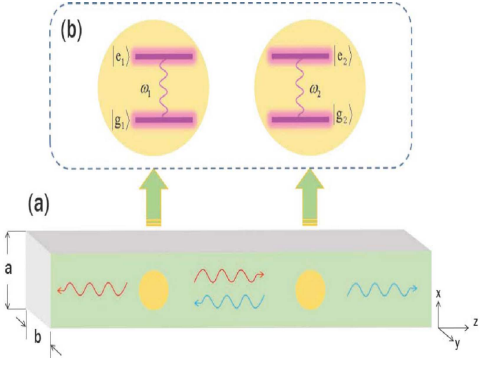}
\caption{(color online) Schematic illustration for an infinite waveguide of
rectangular cross section $A=ab$ (a) coupling to two TLSs (b) located at $%
\vec{r}_{1}=(a/2,b/2,z)$ and $\vec{r}_{2}=(a/2,b/2,z+d)$.}
\label{Fig1.eps}
\end{figure}
The rectangular hollow metallic waveguide made of perfect conductors is
confined in the $x$-$y$ plane with the area $A=ab$ of its cross section,
and translational invariant in the $z$ direction, as shown in Fig.~\ref{Fig1.eps}.
For the convenience of later discussion, we set $a=2b$. The fields in the rectangular
waveguide are classified as transverse magnetic (TM) or transverse electric (TE)
according to whether the electric field or magnetic field transverse to the axial
direction of the guide. Each guiding mode is characterized by three wave
numbers $\{m\pi /a,n\pi /b,k\}$. Its dispersion relation is given by $\omega
_{mnk}$=$\sqrt{\Omega _{mn}^{2}+(ck)^{2}}$, where $\Omega _{mn}=c\sqrt{(m\pi
/a)^{2}+(n\pi /b)^{2}}$ is the cutoff frequency. We note that we only study
the role of the guided modes in this paper and the evanescent modes are not
considered. The free Hamiltonian of the fields reads
\begin{equation}
\hat{H}_{f}=\sum_{j}\int dk\hbar \omega _{jk}\hat{a}_{jk}^{\dagger }\hat{a}_{jk}
\label{2-A3}
\end{equation}%
where $\hat{a}_{jk}^{\dagger }$ ($\hat{a}_{jk}$) is the creation
(annihilation) operator of the $TM_{mn}$ modes. Here, the numbers $(m,n)$
have been replaced with the sequence number $j$, i.e., $j=1,2,3...$. The
reason why only $TM_{mn}$ modes are considered will be given in the
following. Two TLSs, named TLS $1$ and TLS $2$, are separately located
inside the waveguide at positions $\vec{r}_{1}=(a/2,b/2,z_{1})$ and $\vec{r}%
_{2}=(a/2,b/2,z_{2})$, the distance between the TLSs is denoted by $%
d=z_{2}-z_{1}$. We assume the dipoles of TLSs are along the $z$ axis. In
this case, only the $TM_{mn}$ guided modes with odd integer $m$ and $n$ are
interacted with the TLSs. The interaction between the TLSs and the the
electromagnetic field is written as
\begin{equation}
\hat{H}_{int}=\sum_{l=1}^{2}\sum_{j}\int dk\frac{ig_{jl}}{\sqrt{\omega _{jk}}}%
e^{ikz_{l}}\hat{S}_{l}^{-}\hat{a}_{k}^{\dagger }+h.c.  \label{2-A4}
\end{equation}%
in the electric dipole and rotating wave approximations, where $%
g_{jl}=\Omega _{j}\mu _{l}\sin \frac{m\pi }{2}\sin \frac{n\pi }{2}/\sqrt{%
A\hbar \pi \epsilon _{0}}$ and $\mu _{l}$ the magnitude of the dipole of the
$l$-th TLS. We assume that $\mu _{l}$ is real. If the dipoles $\mu _{l}=\mu $%
, the parameter $g_{jl}$ is independent of the subscript $l$ and it becomes
\begin{equation}
g_{j}=\frac{\Omega _{mn}\mu }{\sqrt{\hbar A\pi \epsilon _{0}}}\sin \frac{%
m\pi }{2}\sin \frac{n\pi }{2},  \label{2-A5}
\end{equation}%
where $\epsilon _{0}$ is the permittivity of free space, and $%
j=(1,1),(3,1),(5,1)\cdots $ in the ascending order.

%%%%%%%%%%%%%%%%%%%%%%%%%%%%%%%%%%%%%%%%%%%%%%%%%%%%%%%%%%%%

\section{\label{Sec:3} Time Evolution of the TLSs}

%%%%%%%%%%%%%%%%%%%%%%%%%%%%%%%%%%%%%%%%%%%%%%%%%%%%%%%%%%%%
In the case of a single excitation present in the system, the state vector
of the system can be written as
\begin{equation}
\left\vert \psi (t)\right\rangle =b_{1}\left\vert eg0\right\rangle
+b_{2}\left\vert ge0\right\rangle +\sum_{j}\int dkb_{jk}\hat{a}_{jk}^{\dagger
}\left\vert gg0\right\rangle   \label{2-B1}
\end{equation}%
where $\left\vert 0\right\rangle $ is the vacuum state of the quantum field,
$b_{l}\left( t\right) ,l=1,2$ is the probability amplitude for TLS $l$ being
excited, $b_{jk}\left( t\right) $ the probability amplitude for the
excitation in a mode $k$ of the TM$_{j}$ guided mode. The initial state of
the system is denoted by the amplitudes $b_{1}\left( 0\right) ,b_{2}\left(
0\right) $, $b_{jk}\left( 0\right) =0$. The Schr\"{o}dinger equation results
in the following coupled equation of the amplitudes
\begin{subequations}
\label{2-B2}
\begin{eqnarray}
\dot{b}_{1} &=&-i\omega _{1}b_{1}-\sum_{j}\int dkb_{jk}\frac{g_{j1}}{\sqrt{%
\omega _{jk}}}e^{-ikz_{1}} \\
\dot{b}_{2} &=&-i\omega _{2}b_{2}-\sum_{j}\int dkb_{jk}\frac{g_{j2}}{\sqrt{%
\omega _{jk}}}e^{-ikz_{2}} \\
\dot{b}_{jk} &=&-i\omega _{jk}b_{jk}+\frac{e^{ikz_{1}}}{\sqrt{\omega _{jk}}}%
\left( g_{j1}b_{1}+g_{j2}b_{2}e^{ikd}\right)
\end{eqnarray}%
We introduce three new variables to remove the high-frequency effect
\end{subequations}
\begin{subequations}
\label{2-B3}
\begin{eqnarray}
b_{1}(t) &=&B_{1}(t)e^{-i\omega _{A}t}, \\
b_{2}(t) &=&B_{2}(t)e^{-i\omega _{A}t}, \\
b_{jk}(t) &=&B_{jk}\left( t\right) e^{-i\omega _{jk}t},
\end{eqnarray}%
and define the mean frequency of the TLSs as well as the difference of the
TLSs' frequencies
\end{subequations}
\begin{equation}
\omega _{A}=\frac{\omega _{2}+\omega _{1}}{2},\delta =\frac{\omega
_{1}-\omega _{2}}{2}  \label{2-B4}
\end{equation}%
Then, we formally integrate equation of $B_{jk}\left( t\right) $, which is
later inserted into the equations for $B_{1}\left( t\right) $ and $%
B_{2}\left( t\right) $. The probability amplitudes for one TLS being excited
are determined by two coupled integro-differential equations. Assuming that
the frequency $\omega _{A}$ is far away from the cutoff frequencies $\Omega
_{j}$, we can expand $\omega _{jk}$ around $\omega _{A}$ up to the linear
term
\begin{equation}
\omega _{jk}=\omega _{A}+v_{j}\left( k-k_{j0}\right) ,  \label{2-B5}
\end{equation}%
where the wavelength of the emitted radiation $k_{j0}=\sqrt{\omega
_{A}^{2}-\Omega _{j}^{2}}/c$ is determined by $\omega _{jk_{0}}=\omega _{A}$%
, and the group velocity
\begin{equation}
v_{j}\equiv \frac{d\omega _{jk}}{dk}|_{k=k_{j0}}=\frac{c\sqrt{\omega
_{A}^{2}-\Omega _{j}^{2}}}{\omega _{A}}  \label{2-B6}
\end{equation}%
is different for different TM$_{j}$ guided modes. Integrating over all wave
vectors $k$ gives rise to a linear combination of $\delta \left( t-\tau
-\tau _{j}\right) $ and $\delta \left( t-\tau \right) $, where $\tau
_{j}=d/v_{j}$ is the time that the emitted photon travels from one TLS to
the other TLS in the given transverse mode $j$. The dynamics of two TLSs is
governed by the differential equations\cite%
{DungPRA59,DornPRA66,RistPRA78,GulfPRA12,JingPLA377}
\begin{subequations}
\label{2-B7}
\begin{eqnarray}
\left( \partial _{t}+\Gamma _{1}+i\delta \right) B_{1}(t) &=&-\sum_{j}\gamma
_{j}e^{i\varphi _{j}}B_{2}\left( t_{j}\right) \Theta \left( t_{j}\right)  \\
\left( \partial _{t}+\Gamma _{2}-i\delta \right) B_{2}(t) &=&-\sum_{j}\gamma
_{j}e^{i\varphi _{j}}B_{1}\left( t_{j}\right) \Theta \left( t_{j}\right)
\end{eqnarray}%
where we have defined the phase $\varphi _{j}=k_{j0}d$ due to the distance
between the TLSs, and $\gamma _{j}={g_{j1}g_{j2}\pi}/(v_{j}\omega _{A})$
are caused by the interaction between the TLSs and the vacuum field in a
given transverse mode $j$, $\Theta \left( x\right) $ is the Heaviside unit
step function. The decay rate of the $l$ TLS to all $TM_{j}$ modes is
denoted by $\Gamma _{l}=\sum_{j}\gamma _{lj}$, where $\gamma
_{lj}=g_{jl}^{2}\pi/(v_{j}\omega _{A})$ is the decay rate of the $l$th TLS to
the continuum of the $TM_{j}$ mode, the retard effect has been implied by
the symbol $t_{j}=t-\tau _{j}$. Eqs.(\ref{2-B7}) that the two separate TLSs
are coupled after the time $\min \tau _{j}$ due to the spontaneous emission
from one TLS to the other by the TLSs coupled to the same modes of the
vacuum field.

%%%%%%%%%%%%%%%%%%%%%%%%%%%%%%%%%%%%%%%%%%%%%%%%%%%%%%%%%%%%

\section{\label{Sec:4} Entanglement Dynamics of the TLSs}

%%%%%%%%%%%%%%%%%%%%%%%%%%%%%%%%%%%%%%%%%%%%%%%%%%%%%%%%%%%%

To measure the amount of the entanglement, we use concurrence as the
quantifier\cite{Wootters}. By taking a partial trace over the degrees of
freedom of the waveguide, the density matrix of the two TLSs is of
an X-form in the two-qubit standard basis $\{\left\vert gg\right\rangle
,\left\vert eg\right\rangle ,\left\vert ge\right\rangle ,\left\vert
ee\right\rangle \}$. The concurrence for this type of state can be
calculated easily as
\end{subequations}
\begin{equation}  \label{2-B8}
C(t)=\max (0,2\left\vert B_{1}(t)B_{2}^{\ast }(t)\right\vert )
\end{equation}
for TLSs initially sharing single excitation.

\subsection{single transverse mode}

A TLS in its excited state radiates waves into the continua of the modes
which are resonant with the TLS. If all TLSs' energy separations lie within
the frequency band between $\Omega _{11}$ and $\Omega _{31}$ and are far way
from the cutoff frequencies $\Omega _{11}$ and $\Omega _{31}$, they only
emit photons into the TM$_{11}$ ($j=1$) guided mode. The equations for the
amplitudes of the TLSs read
\begin{subequations}
\label{3-A1}
\begin{eqnarray}
\left( \partial _{t}-i\xi _{1}\right) B_{1}(t) &=&-\alpha _{1}B_{2}\left(
t_{1}\right) \Theta \left( t_{1}\right)  \\
\left( \partial _{t}-i\xi _{2}\right) B_{2}(t) &=&-\alpha _{1}B_{1}\left(
t_{1}\right) \Theta \left( t_{1}\right)
\end{eqnarray}%
\end{subequations}
where $\xi _{1}=i\gamma _{11}-\delta $, $\xi _{2}=i\gamma _{21}+\delta $ and
$\alpha _{1}=\gamma _{1}e^{i\varphi _{1}}$.
By assuming that the TLSs are excited initially and there is no photons in
the field, the Laplace transform of Eqs.(\ref{3-A1}) leads to
\begin{subequations}
\label{3-A2}
\begin{eqnarray}
B_{1}(s) &=&\frac{\left( s-i\xi _{2}\right) B_{1}\left( 0\right) -\alpha
_{1}e^{-s\tau _{1}}B_{2}\left( 0\right) }{\left( s-i\xi _{2}\right) \left(
s-i\xi _{1}\right) -\left( \alpha _{1}e^{-s\tau _{1}}\right) ^{2}}, \\
B_{2}\left( s\right)  &=&\frac{\left( s-i\xi _{1}\right) B_{2}\left(
0\right) -\alpha _{1}e^{-s\tau _{1}}B_{1}\left( 0\right) }{\left( s-i\xi
_{1}\right) \left( s-i\xi _{2}\right) -\left( \alpha _{1}e^{-s\tau
_{1}}\right) ^{2}}.
\end{eqnarray}
\end{subequations}
The integrand in the inverse Laplace transform yields the time-dependent
amplitudes of the TLSs. Defining $t_{1}^{(n)}=t-n\tau _{1}$, the integrand
can be expanded into a power series
\begin{subequations}
\label{3-A3}
\begin{eqnarray}
B_{1}(t) &=&\sum_{n=0}^{\infty }B_{1}\left( 0\right) \Theta \left(
t_{1}^{(2n)}\right) \left[ A_{n}\left( \xi _{2}\right) +B_{n}\left( \xi
_{1}\right) \right]  \\
&&-\sum_{n=0}^{\infty }B_{2}\left( 0\right) \Theta \left(
t_{1}^{(2n+1)}\right) \left[ C_{n}\left( \xi _{2}\right) +C_{n}\left( \xi
_{1}\right) \right] ,  \notag \\
B_{2}(t) &=&\sum_{n=0}^{\infty }B_{2}\left( 0\right) \Theta
\left(
t_{1}^{(2n)}\right) \left[ A_{n}\left( \xi _{1}\right) +B_{n}\left( \xi
_{2}\right) \right]  \\
&&-\sum_{n=0}^{\infty }B_{1}\left( 0\right) \Theta \left(
t_{1}^{(2n+1)}\right) \left[ C_{n}\left( \xi _{1}\right) +C_{n}\left( \xi
_{2}\right) \right] ,   \notag
\end{eqnarray}
\end{subequations}
where the functions are defined as
\begin{subequations}
\label{3-A4}
\begin{eqnarray}
A_{n}\left( \xi _{k}\right)  &=&\lim_{z\rightarrow \xi _{k}}\frac{d^{n-1}}{%
dz^{n-1}}\frac{\left( -i\alpha _{1}\right) ^{2n}e^{izt_{1}^{(2n)}}}{\left(
n-1\right) !\left( z-\xi _{l}\right) ^{n+1}}, \\
B_{n}\left( \xi _{k}\right)  &=&\lim_{z\rightarrow \xi _{k}}\frac{d^{n}}{%
dz^{n}}\frac{\left( -i\alpha _{1}\right) ^{2n}e^{izt_{1}^{(2n)}}}{n!\left(
z-\xi _{l}\right) ^{n}}, \\
C_{n}\left( \xi _{k}\right)  &=&\lim_{z\rightarrow \xi _{k}}\frac{d^{n}}{%
dz^{n}}\frac{\left( -i\gamma _{1}e^{i\varphi _{1}}\right)
^{2n+1}e^{izt_{1}^{(2n+1)}}}{n!\left( z-\xi _{l}\right) ^{n+1}}.
\end{eqnarray}
\end{subequations}
with subscripts $k,l\in \left\{ 1,2\right\} $ and $k\neq l$. The TLSs show
different behavior depending on the retardation times $n\tau _{1}$ required
for light to travel between two TLSs located at a finite distance, the phase
$\varphi _{1}$, the difference $\delta $ of the TLSs' frequencies and the
difference $\gamma _{11}-\gamma _{21}$ of the decaying factors. For $t\in %
\left[ 0,\tau _{1}\right] $, only one term appears
\begin{equation}
B_{l}(t)=B_{l}(0)e^{i\xi _{l}t}.  \label{3-A5}
\end{equation}%
Two TLSs decays as if they are isolated in the waveguide, so the concurrence
decay from $\left\vert B_{1}(0)B_{2}(0)\right\vert $ with a rate $\gamma
_{11}+\gamma _{21}$. As long as $\left( \gamma _{11}+\gamma _{21}\right)
\tau _{1}\gg 1$, two TLSs emit photons independently, the photon travels
along the waveguide for time $\tau _{1}$, then the part toward the other TLS
will be absorbed, and the TLSs is partially excited, later TLSs reemit the
photon again, the whole process of emission and absorption was repeated as
time goes on, however no interference occurs, so the phase $\varphi _{1}$
has no effect on the entanglement. Eq.(\ref{3-A5}) is also the solution of
Eq. (\ref{3-A1}) with $\tau _{1}\rightarrow \infty $. If the TLSs decay
slowly so that $\left( \gamma _{11}+\gamma _{21}\right) \tau _{1}$ is
smaller than or equal to $1$, it is possible for a TLS to be aware of the
other very soon, interference can be produced by multiple reemissions and
reabsorptions of light, which leads to an oscillatory energy exchange
between the TLSs. It can be easily seen that the atomic upper state
population contains two terms for $t\in \left[ \tau _{1},2\tau _{1}\right] $%
\begin{subequations}
\label{3-A6}
\begin{eqnarray*}
B_{1}(t) &=&B_{1}(0)e^{i\xi _{1}t}-B_{2}\left( 0\right) \alpha _{1}\frac{%
e^{i\xi _{2}\left( t-\tau _{1}\right) }-e^{i\xi _{1}\left( t-\tau
_{1}\right) }}{i\left( \xi _{2}-\xi _{1}\right) } \\
B_{2}(t) &=&B_{2}\left( 0\right) e^{i\xi _{2}t}-B_{1}\left( 0\right) \alpha
_{1}\frac{e^{i\xi _{2}\left( t-\tau _{1}\right) }-e^{i\xi _{1}\left( t-\tau
_{1}\right) }}{i\left( \xi _{2}-\xi _{1}\right) }
\end{eqnarray*}
\end{subequations}
The second term in the last equation shows that the TLS is aware of the
other, and presents the absorption and reemission of the other TLS, so
interference is possible. As time goes on, multiple reemissions and
reabsorptions of photons appear which implied in the summation in Eq.(\ref%
{3-A3}), then the phase $\varphi _{1}$ has the influence on the energy
exchange between TLSs.
\begin{figure}[tbp]
\includegraphics[bb=35bp 170bp 343bp 751bp,width=7cm]{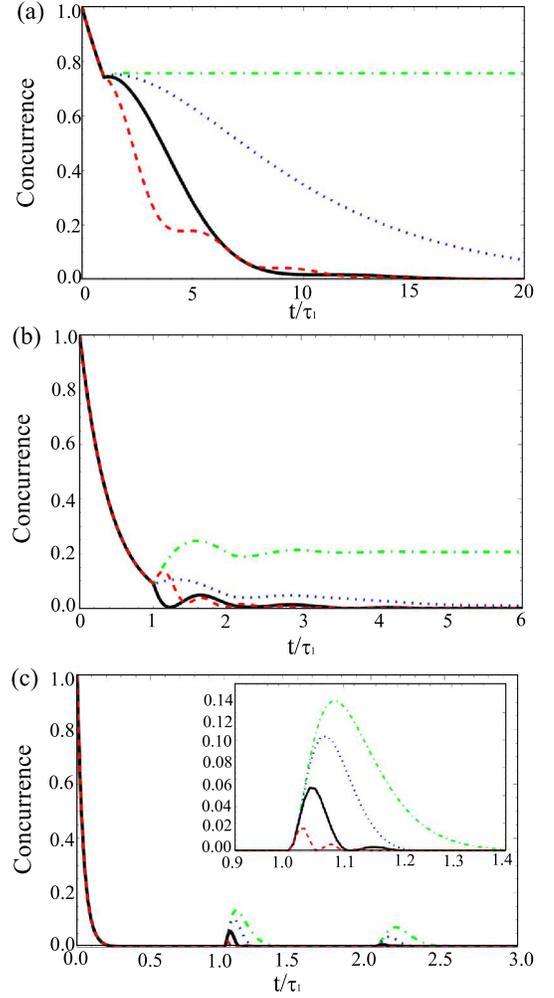}
\caption{(Color online) The concurrence between the TLSs as functions of the
dimensionless time $t/\tau_{1}$ with the TLSs initially in the antisymmetry state $|a\rangle$
for a difference energy separation $\delta=0$ (green dot-dashed line), $\delta=\gamma$ (blue dotted
line), $\delta=2\gamma$ (black solid line), $\delta=5\gamma$ (red dashed
line) in (a) $n=1.0\times 10^{9}$, (b) $n=8.0\times 10^{9}$, (c) $n=8.0\times
10^{10}$. We have set the following parameters: $a=2b$, $\omega_{A}=(\Omega_{11}+\Omega_{31})/2$,
$\gamma _{1}\lambda_{1}/v_{1}=1.5\times 10^{-10}$, $\varphi _{1}=2n\pi$.}\label{fig:2}
\end{figure}

\begin{figure}[tbp]
\includegraphics[bb=3bp 47bp 399bp 179bp,width=9cm]{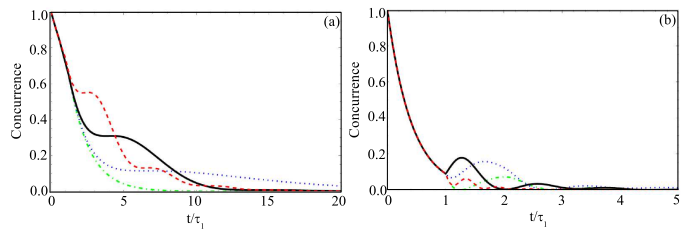}
\caption{(Color online) The concurrence between the TLSs as functions of the
dimensionless time $t/\tau_{1}$ with the TLSs initially in the symmetry state $|s\rangle$
for a difference energy separation $\delta=0$ (green dot-dashed line), $\delta=\gamma$ (blue dotted
line), $\delta=2\gamma$ (black solid line), $\delta=5\gamma$ (red dashed
line) in (a) $n=1.0\times 10^{9}$, (b) $n=8.0\times 10^{9}$. Other parameters are the same as Fig.~\ref{fig:2}.}
\label{fig:3}
\end{figure}

To show the effect of the energy difference $\delta $ of the uncoupled state
$|eg\rangle $ and $|ge\rangle $ on the entanglement dynamics, we consider
the dipoles $\mu _{l}=\mu $, so the damping rates $\gamma _{11}=\gamma
_{12}=\gamma _{1}$. It is known in Ref.~\cite{LijuanQIP} that there is a
dark state for two identical TLSs which is completely isolated and evolves
independently, it can preserve the concurrence, the dark state could be the
symmetry state $|s\rangle =(|eg\rangle +|ge\rangle )/\sqrt{2}$ and the
antisymmetry state $|a\rangle =(|eg\rangle -|ge\rangle )/\sqrt{2}$ could
have a maximum damping rate $2\gamma $ when the phase $\varphi
_{1}=(2n+1)\pi $; the antisymmetry state $|a\rangle $ is the dark state and
the symmetry state $|s\rangle $ could have a maximum damping rate $2\gamma $
when the phase $\varphi _{1}=2n\pi $. In terms of the amplitudes $%
C_{a}\left( t\right) $ and $C_{s}\left( t\right) $ of the antisymmetry and
symmetry state, Eq.(\ref{3-A1}) can be written as
\begin{subequations}
\label{3-A7}
\begin{eqnarray}
\left( \partial _{t}+\gamma _{1}\right) C_{s}\left( t\right)  &=&-i\delta
C_{a}\left( t\right) -\alpha _{1}C_{s}\left( t_{1}\right) \Theta \left(
t_{1}\right)  \\
\left( \partial _{t}+\gamma _{1}\right) C_{a}\left( t\right)  &=&-i\delta
C_{s}\left( t\right) +\alpha _{1}C_{a}\left( t_{1}\right) \Theta \left(
t_{1}\right)
\end{eqnarray}
\end{subequations}
In Fig.~\ref{fig:2}, we plot the concurrence as a function of the time in unit of
$\tau_1$ with the TLSs initially in the antisymmetry state $|a\rangle$ for
$\gamma _{1}\lambda_{1}/v_{1}=1.5\times 10^{-10}$ where the wavelength $\lambda_1k_{10}=2\pi$.
The evolution of $C(t)$ is profoundly affected by the difference $\delta$, phase $\varphi_1$
and delay time $\tau _{1}$, where $\varphi_1$ and $\tau _{1}$ are introduced by the inter-TLS
distance. It can be observed from Fig.~\ref{fig:2}(a) that the evolution of $C(t)$ is independent
of the finite propagating time of the light for the delay time $\tau _{1}\ll \gamma_{1}^{-1}$,
the two TLSs act collectively.  When $\delta=0$ the antisymmetry state $|a\rangle$ is a dark
state which preserve the entanglement among the TLSs; as $\delta$ increases but still smaller
than $2\gamma$, the concurrence decreases monotonous as time increases, as $\delta$ is larger
than $2\gamma$, the concurrence decreases non-monotonically. The dependence of the entanglement
on $\delta $ in Fig.~\ref{fig:2}(a) can be understood by letting $\tau _{1}\rightarrow 0$. In
this case, Eq.(\ref{3-A7}) becomes
\begin{eqnarray*}
\partial _{t}C_{s}\left( t\right)  &=&-i\delta C_{a}\left( t\right) -2\gamma
_{1}C_{s}\left( t\right)  \\
\partial _{t}C_{a}\left( t\right)  &=&-i\delta C_{s}\left( t\right)
\end{eqnarray*}
As long as $\delta \neq 0$, the energy difference $\delta $ of two TLSs
introduces the coupling between state $|s\rangle $ and $|a\rangle $.
Symmetry state $|s\rangle $ is not only coupled to antisymmetry state $%
|a\rangle $ but also coupled to the broad continua of the field, the
coupling of the state $|s\rangle $ to the field introduces the dissipation,
which characterized by the damping rate $2\gamma $. Energy loss occurs when $%
|s\rangle $ is populated. When $2\gamma >\delta $, the loss out of the two
TLSs is the dominant coupling, the initially unoccupied state $|a\rangle $
exchanges energy with state $|s\rangle $, but the energy in state $|s\rangle
$ losses to the field quickly, so it can not be back to state $|a\rangle $,
this is why the concurrence is a monotonically decreasing function of time.
When $2\gamma <\delta $, the population in state $|s\rangle $ can
transferred back to state $|a\rangle $, so the concurrence undergoes
oscillations before decaying to zero. As the inter-TLS separation increases
a little bit to meet $\tau _{1}\sim \gamma _{1}^{-1}$ in Fig.~\ref{fig:2}%
(b), the interference produced by multiple reemissions and reabsorptions of
photon results in an oscillatory entanglement even when $\delta =0$. However
the exchange of population reduces the magnitude of the concurrence.
Panel (c) of Fig.~\ref{fig:2} illustrates the dynamics of entanglement for
a larger inter-TLS distance with $\tau _{1}\gg \gamma _{1}^{-1}$. It can be
observed that at time interval $\left[ 0,\tau _{1}\right] $, each initially
excited TLS emits light to the waveguide, and the entanglement decays
exponentially from unity to zero. The radiation field emitted into the
waveguide returns to the TLSs after $\tau _{1}$, then the entanglement is
created. But, the periodic maxima of the concurrence are in magnitude as
time increases due to the energy loss carried away by the forward-going and
the backward-going waves. Population exchange introduced by the energy
difference $\delta $ further lower the periodic maxima of the concurrence,
however, oscillations can be observed when $\delta >2\gamma $ after time $%
\tau _{1}$. We would like to note that non-vanishing $\delta $ can also raise
the transient behaviors of the concurrence if the TLSs are initially in the
symmetry state $|s\rangle $ with $\varphi _{1}=2n\pi $, as shown in Fig.~\ref{fig:3}.

\subsection{two transverse modes}
As the energy splitting of both TLS increases so that they are much larger than
the cutoff frequency $\Omega _{31}$ and much smaller than $\Omega _{51}$, the
TLSs interact with the field of both TM$_{11}$ and TM$_{31}$ guided modes.
For dipoles $\mu _{l}=\mu $, the equations for the amplitudes of the
symmetry and antisymmetry states reads
\begin{subequations}
\label{3-B1}
\begin{eqnarray}
&& \partial _{t}C_{s}(t)+\Gamma C_{s}(t)+i\delta C_{a}(t)\\
&=&-\alpha _{1}C_{s}\left( t_{1}\right) \Theta \left(
t_{1}\right) -\alpha _{2}C_{s}\left( t_{2}\right) \Theta \left( t_{2}\right)
 \notag \\
&& \partial _{t}C_{a}(t)+\Gamma C_{a}(t)+i\delta C_{s}(t) \\
&=&\alpha _{1}C_{a}\left( t_{1}\right) \Theta \left(
t_{1}\right) +\alpha _{2}C_{a}\left( t_{2}\right) \Theta \left( t_{2}\right)
 \notag
\end{eqnarray}
\end{subequations}
where $\Gamma =\gamma _{1}+\gamma _{2}$ and $\alpha _{j}=\gamma
_{j}e^{i\varphi _{j}}$ ($j=1,2$). The definitions of delay time $\tau _{j}$
and phase $\varphi _{j}$ indicate that $\tau _{j}<\tau _{j+1}$ and $\varphi
_{j}<\varphi _{j+1}$ for a given TLSs' separation. Through an inspection of
Eq.(\ref{3-B1}) for $d=0$, it can be found that the antisymmetry state $%
\left\vert a\right\rangle $ is still a dark state and the symmetry state $%
\left\vert s\right\rangle $ has a maximum damping rate $2\Gamma $. As $d$
increases a little bit but still satisfying $\varphi _{1}=2n\pi $ and the
energy difference $\delta =0$, state $\left\vert a\right\rangle $ is no
longer a dark state, it damps with a damping rate $\gamma _{2}-\text{Re}%
\alpha _{2}$, the width of state $\left\vert s\right\rangle $ becomes $%
2\gamma _{1}+\gamma _{2}+\text{Re}\alpha _{2}$ which is smaller than $%
2\Gamma $, but there is a energy splitting $2\text{Im}\alpha _{2}$ between
the two states. when $\delta \neq 0$, the two states are coupled to each
other, so the oscillation strength is changed from $\Omega =\delta $ in one guiding
mode to $\Omega =\sqrt{\delta ^{2}+\left(\text{Im}\alpha _{2}\right) ^{2}}$ in two
guiding modes, the damping rates of the two states are also changed. As long
as the oscillation strength $\Omega$ is larger than the sum $2\Gamma $, the population
will oscillate obviously when the effect of the phase on the dynamics is more important
than the delay time.
\begin{figure}[tbp]
\includegraphics[bb=0bp 156bp 452bp 455bp,width=9cm]{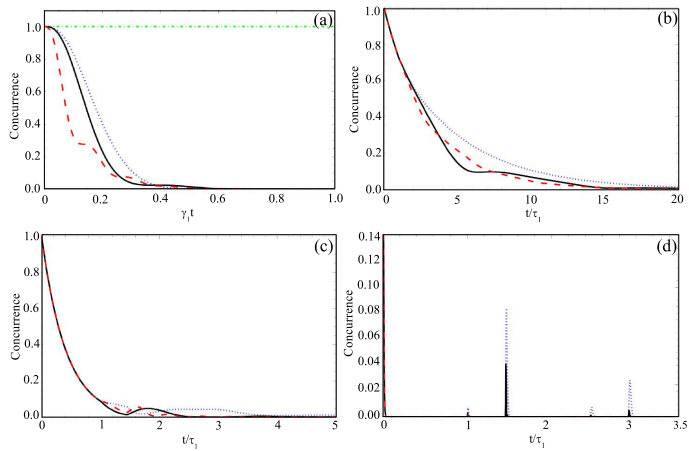}
\caption{(Color online) The concurrence of the TLSs as functions of the
dimensionless time $t/\tau_{1}$ with the TLSs initially in the antisymmetry
state $|a\rangle$ and $a=2b$, $\omega_{A}=(\Omega_{31}+\Omega_{51})/2$,
$\gamma _{1}\lambda_{1}/v_{1}=1.1\times 10^{-11}$, and $\varphi _{1}=2n\pi$.
Panel (a): $n=0$, $\delta=0$ for dash-dotted green line, $\delta=2\gamma_2$ for
dotted blue line, $\delta=2(\gamma_1+\gamma_2)$ for black solid line,
$\delta=5(\gamma_1+\gamma_2)$ for dashed red line. In panels (b)-(d), $\delta=0$ for dotted blue line,
$\delta=1.5\gamma_1+2\gamma_2$ for solid black line, $\delta=5(\gamma_1+\gamma_2)$ for
dashed red line, and the phases are different: $n=3.1\times 10^{9}$ in (b),
$n=2.3\times 10^{10}$ in (c), $n=1.8\times 10^{12}$ in (d).}\label{fig:4}
\end{figure}

\begin{figure}[tbp]
\includegraphics[bb=0bp 70bp 412bp 330bp,width=9cm]{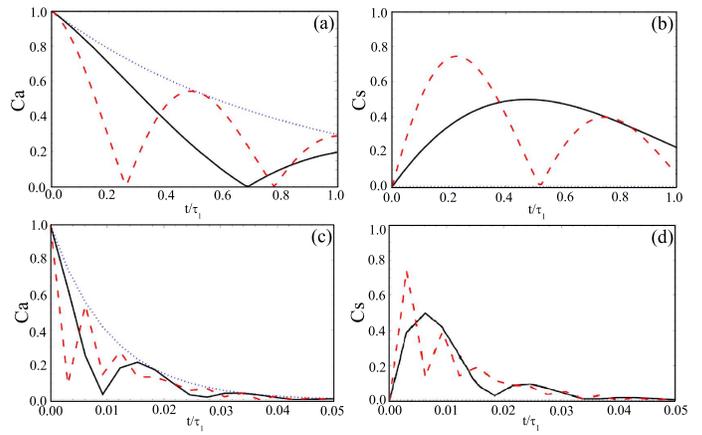}
\caption{(Color online) The populations of the symmetry and antisymmetry states in the time
interval $[0,\tau_{1}]$ of Fig.~\ref{fig:3} (c) and (d). Here, $\delta=0$ for dotted blue line,
$\delta=1.5\gamma_1+2\gamma_2$ for solid black line, $\delta=5(\gamma_1+\gamma_2)$ for
dashed red line and $\varphi _{1}=2n\pi$:
$n=2.3\times 10^{10}$ in (a, b), $n=1.8\times 10^{12}$ in (c, d).}\label{fig:5}
\end{figure}

In Fig.~\ref{fig:4}, we have plotted the concurrence between the TLSs as a function of the
dimensionless time $t/\tau_{1}$ with the TLSs initially in the antisymmetry state $|a\rangle$.
Panel (a) shows the entanglement dynamics when the inter-TLS distance $d=0$ with different
$\delta$, it can be found that the concurrence remain its initial value when $\delta=0$, however,
as $\delta$ increases until $2\Gamma$, the faster the population changes between the two states,
the faster the concurrence decays. As $\delta$ increases further, i.e. more than $2\Gamma$, there
is an oscillation. In panel (b), the delay time $\tau_2\ll \gamma_2^{-1}$, which
means there is no delay in the absorption of the energy by another TLS in both $TM_{11}$ and $TM_{31}$ modes.
The antisymmetry state interacts with the field in $TM_{31}$ guiding mode due to the phase
$\varphi_2\neq 2n\pi$, the concurrence undergoes an exponential decay when $\delta=0$. Although
the concurrence is further decreased by the increasing of $\delta$, its evolution deviates from
the exponential decay. The dotted blue line in panel (c) exhibits a behavior different from
that in panel (b), which indicates that phases and delay times play an equal role. In the interval
$[0,\tau_2]$, the concurrence exponentially decay with a rate $\Gamma$ up to time $\tau_1$.
After this, it still decreases but deviates from the exponential decay, which means that the
phase $\varphi_1$  begins to have an effect until time $t=\tau_2$. After time $\tau_2$, the
dynamics can be dramatically affected by the phases $\varphi_j$, delay times $\tau_j$ when
$\delta=0$. However, the energy difference $\delta$ can increase the entanglement. We note
that in the interval $[0,\tau_1]$, the exponential decay of the concurrence is independent
of $\delta$, but it is possible for the population of state $|a\rangle$ to present an
oscillating behavior as shown in Fig.~\ref{fig:5} since the amplitudes obey the following
equation
\begin{subequations}
\label{3-B2}
\begin{eqnarray}
&& \partial _{t}C_{s}(t)+\Gamma C_{s}(t)+i\delta C_{a}(t)= 0 \\
&& \partial _{t}C_{a}(t)+\Gamma C_{a}(t)+i\delta C_{s}(t) = 0
\end{eqnarray}
\end{subequations}
This equation can also explain the exponential decay of panel (d) of Fig.~\ref{fig:4} in the time
interval $[0,\tau_1]$. Actually, Eq.~(\ref{3-B2}) also describes the dynamics of TLSs when the
delay time $\tau_1\rightarrow \infty$, it indicates that two TLSs emit photons independently, and
the emitted photon travels along the waveguide and is never absorbed by the TLSs. For two TLSs far
apart as shown in Fig.~\ref{fig:4}(d), the phase $\varphi_j$ do not make an sense, and the increase
of $\delta$ lowers the magnitude of the concurrence.Although the emitted photon could be absorbed
by the TLS resulting the birth of entanglement, but the revival time becomes $p\tau_1+q\tau_2$
with $p, q$ are integer, which can be obtained by performing the Laplace transformation on
Eq.~(\ref{2-B7}) with $j=1,2$.

%%%%%%%%%%%%%%%%%%%%%%%%%%%%%%%%%%%%%%%%%%%%%%%%%%%%%%%%%%%%

\section{\label{Sec:5}conclusion}

%%%%%%%%%%%%%%%%%%%%%%%%%%%%%%%%%%%%%%%%%%%%%%%%%%%%%%%%%%%%
We have studied the entanglement dynamics of two distinguishable TLSs characterized by
energy difference $\delta$ located inside a rectangular hollow metallic waveguide of transverse
dimensions $a$ and $b$. The effects of energy difference $\delta$ and the inter-TLS distance
on the time evolution of the concurrence of the TLSs are examined in the single excitation
subspace when the energy separation of the TLS is far away from the cutoff frequencies of the
transverse mode. The inter-TLS distance induces phase factors and delay times in the delay
differential equations. The energy difference introduces the coupling between the symmetry
and antisymmetry state. For the inter-TLS distance $d=0$, the entanglement can be
trapped in the antisymmetry state when $\delta=0$ since the antisymmetry state is decoupled
with the guiding mode, however, the population exchange induced by non-vanishing $\delta$
decreases the entanglement from one to zero. As $d$ increases to satisfy $\max\{\tau _{j}\}\ll\gamma_j^{-1}$,
two TLSs behave collectively. It is well known that a change of phase leads to an enhanced or
inhibited exponential decay of the concurrence, however, $\delta$ makes the dynamics of the
concurrence deviating from the exponential decay. As $d$ increases further so that
$\max\{\tau _{j}\}\approx\gamma_j^{-1}$, although the interference produced by multiple
reemissions and reabsorptions of photon results in the dynamic behavior of the entanglement
deviating from the exponential decay, non-zero $\delta$ can raise the entanglement in transient
as time increases. When $\tau _{j}\gg \gamma_j^{-1}$, an increasing of $\delta$ only lower
the entanglement.

\begin{acknowledgments}
This work was supported by NSFC Grants No. 11975095, No. 12075082, No.
11935006.
\end{acknowledgments}

\end{document}